\documentclass[11pt,a4paper,reqno]{article}
\usepackage{comment}
\usepackage[utf8]{inputenc}
\usepackage[T1]{fontenc}
\usepackage{mathtools, blkarray, booktabs}
\usepackage{amssymb}
\usepackage{mathtools}
\usepackage{latexsym}
\usepackage{amsmath}
\usepackage{graphicx}
\usepackage{relsize}
\usepackage{amsthm}
\usepackage{empheq}
\usepackage{wasysym}

\usepackage{bm}
\usepackage{booktabs}
\usepackage[dvipsnames]{xcolor}
\usepackage{tikz,pgfplots,lipsum,lmodern}
\usepackage{enumitem}
\usepackage[most]{tcolorbox}

\newcommand{\diagdots}[3][-25]{%
  \rotatebox{#1}{\makebox[0pt]{\makebox[#2]{\xleaders\hbox{$\cdot$\hskip#3}\hfill\kern0pt}}}%
}

\usepackage[margin=2.5cm]{geometry}

 \usepackage[section]{placeins}

\numberwithin{equation}{section}

\theoremstyle{definition}
\newtheorem{theorem}{Theorem}[section]

\newtheorem{remark}[theorem]{Remark}
\newtheorem{lemma}[theorem]{Lemma}

\usepackage{calrsfs}

\usepackage{calrsfs}
\DeclareMathAlphabet{\pazocal}{OMS}{zplm}{m}{n}

\newcommand{\fq}{\mathbb{F}_{q}}

\newcommand{\fqQ}{\mathbb{F}_{q}[Q_{4n}]}

\newcommand{\Mat}{\textnormal{Mat}}

\newcommand*{\myproofname}{Proof of the claim}

\makeatletter
\renewcommand*\env@matrix[1][*\c@MaxMatrixCols c]{%
  \hskip -\arraycolsep
  \let\@ifnextchar\new@ifnextchar
  \array{#1}}
\makeatother

\usepackage{hyperref}

\title{
\textbf{Block components of generalized quaternion group codes}}

\date{}                    

\usepackage{setspace}

 \allowdisplaybreaks
\usepackage{longtable}
\usepackage{multirow}

\usepackage{authblk}
\author[1]{Nadja Willenborg}

\affil[1]{University of St.Gallen, Switzerland}

\begin{document}

\maketitle
	
\thispagestyle{empty}
	
\begin{abstract}
Codes in the generalized quaternion group algebra $\fqQ$ are considered. Restricting to $\textnormal{char}\fq \nmid 4n$ the structure of an arbitrary code $C \subseteq \fqQ$ is described via the Wedderburn decomposition. Moreover it is known that in this case every code $C \subseteq \fqQ$ has a generating idempotent $\lambda \in \fqQ$. Given the generating idempotent of a code $C$ we determine the different components in its decomposition $C \cong \bigoplus_{j=1}^{r+s}C_j \oplus \bigoplus_{i=1}^{k+t}C'_{i}.$ Afterwards we apply this result to describe the blocks of codes induced by cyclic group codes.
\end{abstract}

\section{Introduction}
The primary motivation of this chapter is rooted in physics where the theory of motion control and analysis are related to quaternion groups. It is also known that lattices in quaternion algebras yield space-time codes that achieve high spectral efficiency on wireless channels with two transmit antennas, see \cite{berhuy2013introduction}. For an overview on proposed post-quantum cryptosystems and digital signature schemes, see for example \cite{galbraith2018computational}, for applications in quantum computation \cite{parzanchevski2018super}. Understanding these applications serves the desire to dive into structured areas of noncommutative groups, to finally get generic group algebra cyclotomically, see \cite{chlouveraki2009blocks} for more details.
In fact, this perspective allows us to identify and understand the block decomposition of codes $C \subseteq \fqQ$ which is required when transforming these codes to lifted product codes as done in \cite{willenborg2023dihedral}.

Informally, the generalized quaternion group $Q_{4n}$\footnote{The cyclic group of order $2n$ is a commutative normal subgroup of $Q_{4n}$ of index $2$, and so $Q_{4n}$ is solvable.} of order $4n$ considered in \cite{gao2021idempotents} is generated by a rotation $x$ by an angle of $\frac{2\pi}{n}$, an arbitrary reflection $y$ and defining relations
$$x^{2n}=1, y^2 = x^n, yxy^{-1}=x^{-1}.$$
Throughout, let the characteristic of $\fq$ be coprime to $4n$. In this case every vector space $C \subseteq \fqQ$ has a generating idempotent, i.e., an element $\lambda \in \fqQ$ such that $\lambda^2 =\lambda$, see \cite{vedenev2020relationship}. It is known from \cite{martinez2015structure, vedenev2020relationship} that for a monic polynomial $f \in \fq[x]/(x^{2n}-1)$ which divides $x^{2n}-1$, the generating idempotent of the ideal $(f(x))\fq [x]/(x^{2n}-1)$ is 
\begin{equation} \label{eq:genidem}
    e_{g}(x)= - \frac{[(g(x)^{*})']^{*}}{2n}\cdot \frac{x^{2n}-1}{g(x)},
\end{equation}
where $g(x)= \tfrac{x^{2n}-1}{f(x)}$. Moreover it is shown in \cite{vedenev2020relationship} that for $g \in \fq[x]/(x^{2n}-1)$, a monic divisor of $x^{2n}-1$ with $g=g_{1} \cdot g_{2}$, it holds
\begin{equation} \label{eq:idemprop}
    e_{g}(x)=e_{g_{1}}(x)+e_{g_{2}}(x).
\end{equation}
Recall that an idempotent is said to be irreducible, if it cannot be decomposed into a sum of several nonzero idempotents. Moreover an idempotent is said to be central if it belongs to the center of $\fqQ$.

For any monic polynomial $g \in \fq[x]$ with $g(0)=a_{0} \neq 0$, the reciprocal polynomial is defined as $g^{*}(x) = a_{0}^{-1}x^{\textnormal{deg}(g)}g(x^{-1})$, see \cite{gao2021idempotents}. We say that $g$ is selfreciprocal if $g =g^{*}$. In particular $g$ and $g^{*}$ have the same roots in its splitting field.

Suppose that $x^n -1$ and $x^n +1 \in \fq[x]$ split into monic irreducible factors as follows:
\begin{align*}
    x^{n}-1&=f_{1}(x)f_{2}(x) \cdots f_{r}(x)f_{r+1}(x)f_{r+1}^{*}(x) \cdots f_{r+s}(x)f_{r+s}^{*}(x),\\
    x^{n}+1 &= g_{1}(x)g_{2}(x) \cdots g_{t}(x)g_{t+1}(x)g_{t+1}^{*}(x) \cdots g_{t+k}(x)g_{t+k}^{*}(x),
\end{align*}
where $f_{j}(x) = f_{j}^{*}(x)$ for $1 \le j \le r$ and $g_{i}(x) = g_{i}^{*}(x)$ for $1 \le i \le t.$ Moreover, $f_{1}(x)=x-1, f_{2}(x)=x+1$ if $n$ is even; and $f_{1}(x)=x-1, g_{1}(x)=x+1$ if $n$ is odd.
Here $\alpha_{j}$ denotes a root of $f_{j}$ and $\beta_{i}$ denotes a root of $g_{i}$. We set $$\delta(n) = \begin{cases}
        1 \quad &\textnormal{ if $n$ is odd,} \\
   2 \quad &\textnormal{ if $n$ is even,}
\end{cases} \quad \mu(n)= \begin{cases}
        1 \quad &\textnormal{ if $n$ is odd,} \\
   0 \quad &\textnormal{ if $n$ is even.}
\end{cases}$$

Moreover we define
$$
F_{j}:=  \begin{cases}
        \fq \quad & 1 \le j \le \delta(n), \\
   \fq[\alpha_{j}+\alpha_{j}^{-1}] \quad & \delta(n) +1 \le j\le r,
   \\
    \fq[\alpha_{j}] \quad & r+1 \le j \le r+s,
\end{cases} \, L_{i}:=  \begin{cases}
        \fq \quad & 1 \le i \le \mu(n), \\
   \fq[\beta_{i}+\beta_{i}^{-1}] \quad & \mu(n) +1 \le i\le t,
   \\
    \fq[\beta_{i}] \quad & t+1 \le i \le t+k,
\end{cases}
$$
and
$$I_{j}(x,y):= \begin{cases}
        \{ax,by:a,b \in \fq\} \quad & 1 \le j \le \delta(n), \\
   \left\{\begin{pmatrix}
ay & -ax \\
by & -bx
\end{pmatrix}: a,b \in F_{j} \right\} \quad & \delta(n) +1 \le j \le r+s,
\end{cases}$$
$$J_{i}(x,y):= \begin{cases}
        \{ax,by:a,b \in \fq\} \quad & 1 \le i \le \mu(n), \\
   \left\{\begin{pmatrix}
ay & -ax \\
by & -bx
\end{pmatrix}: a,b \in L_{i} \right\} \quad & \mu(n) +1 \le i \le t+k.
\end{cases}$$

Note that $I_{1}(1,0)= \fq \oplus 0, I_{1}(0,1)= 0 \oplus \fq$. Thus for $1 \le j \le \delta(n)$ it suffices to consider $I_{j}(0,1)$ and $I_{j}(1,0)$. Similar observations are useful for $1 \le i \le \mu(n)$. For $\delta(n)+1 \le j\le r+s, \mu(n)+1 \le i \le k+t$ we consider $I_{j}(0,1)$ and $I_{j}(q_j,1)$, respectively $J_{i}(0,1), J_{i}(q_{i},1)$ where $q_j \in F_{j}$, respectively $q_i \in L_{i}$.

By Maschke's theorem the group algebra $\fqQ$ is semisimple and hence isomorphic to a direct sum of matrix algebras over suitable extension fields, realized in \cite{gao2021idempotents} by the map $\rho$,
$$\rho: \fq[Q_{4n}] \rightarrow \left( \bigoplus_{j=1}^{r+s}A_{j}\right) \oplus \left( \bigoplus_{i=1}^{t+k}B_{i}\right),$$
which is defined by families of $\fq$-algebra homomorphisms $(\tau_{j})_{1 \le j \le r+s}$, $(\eta_{i})_{1 \le i \le k+t}$ such that
\begin{align*}
    A_{j}&= \begin{cases}
      \fq \oplus \fq \quad & j \le \delta(n), \\
  \Mat_{2}(\fq[\alpha_{j}+\alpha_{j}^{-1}]) \quad & \delta(n)+1 \le j \le r,\\
  \Mat_{2}(\fq[\alpha_{j}]) \quad & r+1 \le j \le r+s,
\end{cases}\\
B_{i}&= \begin{cases}
      \fq \oplus \fq \quad & i \le \mu(n), \\
  \Mat_{2}(\fq[\beta_{i}+\beta_{i}^{-1}]) \quad & \mu(n)+1 \le i\le t,\\
  \Mat_{2}(\fq[\beta_{i}]) \quad & t+1 \le i \le t+k.
\end{cases}
\end{align*}
As in \cite{vedenev2020codes} one can observe that the inverse $\rho^{-1}$ is well-defined and each direct summand $(A_{j})_{1 \le j\le r+s},\\ (B_{i})_{1 \le i\le t+k}$ has an identity element $\textnormal{id}_{A_{j}}$, respectively $\textnormal{id}_{B_{i}}$, such that the central irreducible idempotents $(e_{j})_{1\le j \le r+s},(e_{i}')_{1 \le i \le t+k}$ are given by
\begin{equation}\label{eq:idemj}
    e_{j}=\rho^{-1}(0, \dots, 0,\textnormal{id}_{A_{j}}, 0, \dots, 0), \, 1 \le j\le r+s,
\end{equation}
\begin{equation}\label{eq:idemi}
    e_{i}'=\rho^{-1}(0, \dots, 0,\textnormal{id}_{B_{i}}, 0, \dots, 0), \, 1 \le i\le k+t.
\end{equation}
Moreover, similarly as in \cite{vedenev2020relationship}, the structure of an arbitrary code $ C \subseteq \fq [Q_{4n}]$ can be described via the decomposition $C \cong \bigoplus_{j=1}^{r+s}C_{j} \oplus \bigoplus_{i=1}^{t+k}C'_{i}$, where
$$C_{j}= \begin{cases}
      A_{j} \quad & j \in S_{1}, \\
   I_{j}(-q_{j},1) \quad & j \in S_{2} \setminus\{1, \delta(n)\},\\
   I_{j}(0,1) \quad & j \in S_{2}\cap\{1, \delta(n)\},\\
   I_{j}(1,0) \quad & j \in S_3,\\
   0 \quad & j \notin S_1 \cup  S_{2} \cup S_3,
\end{cases} \quad C'_{i}= \begin{cases}
      B_i \quad & i\in S_{4}, \\
   J_{i}(-q'_{i},1)\quad & i \in S_{5} \setminus\{1, \mu(n)\},\\
   J_{i}(0,1) \quad & i \in S_{5}\cap\{1, \mu(n)\},\\
   J_{i}(1,0) \quad & i \in S_6,\\
   0 \quad & i \notin S_4 \cup  S_{5} \cup S_6,
\end{cases}$$
  $S_{1}, S_{2},S_{3} \subseteq [r+s], S_{4}, S_{5},S_{6} \subseteq [k+t]$ are pairwise disjoint and $(q_{j})_{j \in S_{2}} \subseteq F_{j}$, $(q'_{i})_{i \in S_{5}}\subseteq L_{i}$.
\section{Main result}
\begin{theorem} \label{thm:main}
    Let $\gcd(4n,q)=1$ and consider an arbitrary code $C \subseteq \fqQ$ generated by its idempotent $\lambda \in \fqQ$. Suppose that
    \begin{align*}
        \phi_{j}&=\lambda e_{f_{j}}(x), \, 1 \le j \le r+s, \quad &\Tilde{\phi}_{j}=\lambda e_{f_{j}^{*}}(x), \, r+1 \le j \le r+s, \\
        \psi_{i} &= \lambda e_{g_{i}}(x), \, 1 \le i \le k+t, \quad &\Tilde{\psi}_{i} = \lambda e_{g_{i}^{*}}(x), \, t+1 \le i \le k+t,
    \end{align*}
    and let $\rho (C)=\displaystyle \bigoplus_{j=1}^{r+s}C_{i} \oplus \bigoplus_{i=1}^{k+t}C_{i}^{'}$ be the code decomposition. Then the code blocks $(C_{j})_{1 \le j \le r+s},(C_{i}')_{1 \le i \le t+k}$ are determined by the elements $\phi_{j}, \Tilde{\phi}_{j}, \psi_{i},\Tilde{\psi}_{i}, 1 \le j \le r+s, 1 \le i \le t+k$ as follows:
    \begin{enumerate}[label=(\roman*)]
     \item For $1 \le j \le \delta(n)$
        $$C_{j}= \begin{cases}
   I_{j}(1,0)\quad & \textnormal{if $\phi_{j}=\tfrac{1+y}{2}e_{f_{j}}(x)$,} \\
  I_{j}(0,1) \quad & \textnormal{if $\phi_{j}=\tfrac{1-y}{2}e_{f_{j}}(x)$,} \\
   0 \quad & \textnormal{if $\phi_{j}=0$,}\\
    A_{j} \quad & \textnormal{if $\phi_{j}=e_{f_{j}}(x)$.}
\end{cases}$$
\item For $\delta(n) +1 \le j \le r$
$$C_{j}= \begin{cases}
   0\quad & \textnormal{if $\phi_{j}=0$,} \\
    A_{j} \quad & \textnormal{if $\phi_{j}=e_{f_{j}}(x)$.}
\end{cases}$$
\item For $r+1 \le j \le r+s$ and $u_j, \Tilde{u_j}, v_j, \Tilde{v_j} \in \fq[x]$ such that $$\phi_{j}=u_{j}(x)+yv_{j}(x), \quad \tilde{\phi_{j}}=\tilde{u_j}(x)+y\tilde{v_j}(x),$$ we have
  \begin{align*}
      C_{j}= \begin{cases}
   I_{j}(\Tilde{v_{j}}(\alpha_{j}^{-1}),-u_{j}(\alpha_{j}))\quad & \textnormal{if $(\tilde{v_j}(\alpha_{j}^{-1}),u_{j}(\alpha_{j}))\neq (0,0)$,} \\
  I_{j}(\Tilde{u_{j}}(\alpha_{j}^{-1}),-v_{j}(\alpha_{j})) \quad & \textnormal{if $\tilde{v_j}(\alpha_{j}^{-1})=u_{j}(\alpha_{j})=0 \wedge (\tilde{u_j}(\alpha_{j}^{-1}),v_{j}(\alpha_{j}))\neq (0,0)$,} \\
   0 \quad & \textnormal{if $\tilde{u_j}(\alpha_{j}^{-1})=v_{j}(\alpha_{j})=\tilde{v_j}(\alpha_{j}^{-1})=u_{j}(\alpha_{j})=0$,}\\
    A_{j} \quad & \textnormal{if $\tilde{v_j}=v_j=0 \wedge u_{j}(x)=e_{f_{j}}(x) \wedge \tilde{u_{j}}(x)=e_{f_{j}}(x)$.}
\end{cases}
  \end{align*}
  \item For $1 \le i \le \mu(n)$ and $q \equiv 1 \pmod{4}$
        \begin{align*}
            C'_{i}= \begin{cases}
   J_{i}(1,0)\quad & \textnormal{if $\psi_{i}=-\left(\tfrac{e_{g_{i}}(x)-\sqrt{-1}y}{2}\right)$,} \\
  J_{i}(0,1) \quad & \textnormal{if $\psi_{i}=-\left(\tfrac{e_{g_{i}}(x)+\sqrt{-1}y}{2}\right)$,} \\
   0 \quad & \textnormal{if $\psi_{i}=0$,}\\
    B_{i} \quad & \textnormal{if $\psi_{i}=-e_{g_{i}}(x)$.}
\end{cases}
        \end{align*}
        For $1 \le i \le \mu(n)$ and $q \equiv 3 \pmod{4}$
        \begin{align*}
            C'_{i}= \begin{cases}
   0 \quad & \textnormal{if $\psi_{i}=0$,}\\
    B_{i} \quad & \textnormal{if $\psi_{i}=-e_{g_{i}}(x)$.}
\end{cases}
        \end{align*}
\item For $\mu(n) +1 \le i \le t$
$$C'_{i}= \begin{cases}
   0\quad & \textnormal{if $\psi_{i}=0$,} \\
    B_{i} \quad & \textnormal{if $\psi_{i}=-e_{g_{i}}(x)$.}
\end{cases}$$
\item For $t+1 \le i \le t+k$ and $u_i, \Tilde{u_i}, v_i, \Tilde{v_i} \in \fq[x]$ such that $$\psi_{i}=u_{i}(x)+yv_{i}(x), \quad \tilde{\psi_{i}}=\tilde{u_i}(x)+y\tilde{v_i}(x),$$
we have
\begin{align*}
    C'_{i}= \begin{cases}
   J_{i}(\Tilde{v_{i}}(\beta_{i}^{-1}),-u_{i}(\beta_{i}))\quad & \textnormal{if $(\tilde{v_i}(\beta_{i}^{-1}),u_{i}(\beta_{i}))\neq (0,0)$,} \\
  J_{i}(\Tilde{u_{i}}(\beta_{i}^{-1}),-v_{i}(\beta_{i})) \quad & \textnormal{if $\tilde{v_i}(\beta_{i}^{-1})=u_{i}(\beta_{i})=0 \wedge (\tilde{u_i}(\beta_{i}^{-1}),v_{i}(\beta_{i}))\neq (0,0)$,} \\
   0 \quad & \textnormal{if $\tilde{u_i}(\beta_{i}^{-1})=v_{i}(\beta_{i})=\tilde{v_i}(\beta_{i}^{-1})=u_{i}(\beta_{i})=0$,}\\
    B_{i} \quad & \textnormal{if $\tilde{v_i}=v_i=0 \wedge u_{i}(x)=e_{g_{i}}(x) \wedge \tilde{u_{i}}(x)=e_{g_{i}}(x)$.}
\end{cases}
\end{align*}
    \end{enumerate}
\begin{proof}
Let $\delta_{ij}$ be the Kronecker delta, i.e., $\delta_{ij}=1$ if $i=j$ and $\delta_{ij}=0$ if $i \neq j.$
    Using the central irreducible idempotents obtained in \cite{gao2021idempotents}, Theorem 3.2 and Theorem 3.8, together with \eqref{eq:idemj} and \eqref{eq:idemi} we have
    $$\sum_{j=1}^{r}e_{f_{j}}+\sum_{j=r+1}^{r+s}(e_{f_{j}}+e_{f_{j}^{*}})-\sum_{i=1}^{t}e_{g_{i}}-\sum_{i=t+1}^{t+k}(e_{g_{i}}+e_{g_{i}^{*}})=1.$$ Let
    $$\rho(\lambda)=\displaystyle \bigoplus_{j=1}^{r+s}\lambda_{j} \oplus \bigoplus_{i=1}^{t+k}\lambda'_{i}, \quad \lambda_{j} \in A_{j}, \lambda'_{i} \in B_{i},$$ giving for $1 \le j \le r$ $$\rho(\phi_{j}) = \displaystyle\bigoplus_{i=1}^{r+s}\lambda_{j}\delta_{ij}$$ and for $r+1 \le j \le r+s$ 
    $$\rho(\phi_{j}+\tilde{\phi_{j}})= \displaystyle\bigoplus_{i=1}^{r+s}\lambda_{j}\delta_{ij}.$$ Similarly for $1 \le i \le t$ $$\rho(\psi_{i}) = \displaystyle\bigoplus_{j=1}^{t+k}\lambda'_{i}\delta_{ij}$$ and for $t+1 \le i \le t+k$
    $$\rho(\psi_{i}+\tilde{\psi_{i}})= \displaystyle\bigoplus_{j=1}^{t+k}\lambda'_{i}\delta_{ij}.$$ Thus the elements $(\lambda_{j})_{1 \le j\le r+s}, (\lambda'_{i})_{1\le i \le t+k}$ can be calculated from $\phi_{j},\tilde{\phi_{j}}, \psi_{i}, \tilde{\psi_{i}}.$ It follows from $$\rho(C) = \rho(\fqQ \lambda)=\displaystyle\bigoplus_{j=1}^{r+s}A_{j}\lambda_{j} \oplus \bigoplus_{i=1}^{t+k}B_{i}\lambda'_{i}$$ that $C_{j}=A_{j}\lambda_{j}$, respectively $C'_{i}=B_{i}\lambda'_{i}.$

    Cases $(i),(ii),(iv)$ and $(v)$ can easily be deduced from $$\rho(\phi_{j})=\bigoplus_{j=1}^{r+s}\lambda_{j}\delta_{ij}, \quad \rho(\psi_{i})=\bigoplus_{i=1}^{t+k}\lambda'_{i}\delta_{ij},$$
    and Theorem 3.2, Theorem 3.8 from \cite{gao2021idempotents} where the central irreducible idempotents covering the possible values of $(\phi_{j})_{1 \le j \le r}, (\psi_{i})_{1 \le i \le t}$ are explicitly given. We exemplarily give arguments proving case $(vi)$ which works similar as case $(iii)$ If $$\psi_{i}=u_{i}(x)+yv_{i}(x), \quad \tilde{\psi_{i}}=\tilde{u_{i}}(x)+y\tilde{v_{i}}(x),$$
    then
    $$\lambda'_{i}\begin{pmatrix}
        1&0\\
        0&0
    \end{pmatrix}=\begin{pmatrix}
        u_i(\beta_{i})&0\\
        v_{i}(\beta_{i})&0
    \end{pmatrix}, \quad \lambda'_{i}\begin{pmatrix}
        0&0\\
        0&1
    \end{pmatrix}=\begin{pmatrix}
        0& \tilde{v_i}(\beta_{i})\\
       0& \tilde{u_i}(\beta_{i})
    \end{pmatrix},$$
   and hence
   $$\lambda'_{i}= \begin{pmatrix}
       u_{i}(\beta_{i})& \tilde{v_i}(\beta_{i}^{-1})\\
        v_{i}(\beta_{i})& \tilde{u_i}(\beta_{i}^{-1})
   \end{pmatrix}.$$
   Obviously, if $\psi_i =\tilde{\psi_{i}}=0,$ then $\lambda'_{i}=0$ and hence $C'_{i}=0.$ If $\psi_{i}=e_{g_{i}}(x)$ and $\tilde{\psi_{i}}=e_{g_{i}^{*}}(x),$ then $\lambda'_{i}=I_{2}$ and $C'_{i}=B_i$. 
   
   Moreover, since $\lambda'_{i}$ is an idempotent by defintion, it follows from the linear independence of the rows of the matrix $\lambda'_{i}$ that $\lambda'_{i}= I_{2}$. Therefore, if $\psi_i \neq 0, \tilde{\psi_i}\neq 0, \psi_i \neq e_{g_{i}}(x), \tilde{\psi_{i}} \neq e_{g_{i}^{*}}(x),$ we have
   $$C'_{i}=\begin{cases}
        J_{i}(\tilde{v_i}(\beta_{i}^{-1}),-u_{i}(\beta_{i}))\quad & \textnormal{if $(\tilde{v_i}(\beta_{i}^{-1}),u_{i}(\beta_{i})) \neq (0,0)$,} \\
    J_{i}(\tilde{u_i}(\beta_{i}^{-1}),-v_{i}(\beta_{i}))\quad & \textnormal{if $(\tilde{u_i}(\beta_{i}^{-1}),v_{i}(\beta_{i})) \neq (0,0)$,} \\
   0 \quad & \textnormal{if $(\tilde{v_i}(\beta_{i}^{-1}),u_{i}(\beta_{i}))=(\tilde{u_i}(\beta_{i}^{-1}),v_{i}(\beta_{i})) = (0,0)$.}\\
   \end{cases}$$
\end{proof}
\end{theorem}
\begin{remark}
    For $r+1 \le j \le r+s$ with $C_{j}=A_j$ we require $\phi_{j}=e_{f_{j}}(x), \tilde{\phi_{j}}=e_{\tilde{f_{j}}}(x)$. Similarly, for $t+1 \le i \le t+k$ with $C_{i}'=B_{i}$ we require $\psi_{i}=e_{g_{i}}(x), \tilde{\psi_{i}}=e_{\tilde{g_{i}}}(x)$.
\end{remark}
\section{Application to induced codes}
In the following we apply Theorem \ref{thm:main} to obtain the structure of codes induced by cyclic groups. Let $\ell$ be a positive integer divisor of $2n$. Consider the cyclic group algebra $\fq[\hat{x}]/(\hat{x}^{\frac{2n}{\ell}}-1)$ and let
$$\hat{x}^{\frac{2n}{\ell}}-1=\left( \prod_{j=1}^{\hat{r}}\hat{f}_{j}(\hat{x}) \right)\left( \prod_{j=\hat{r}+1}^{\hat{r}+\hat{s}}\hat{f}_{j}(\hat{x})\hat{f}_{j}^{*}(\hat{x}) \right)\left( \prod_{i=1}^{\hat{t}}\hat{g}_{i}(\hat{x}) \right)\left( \prod_{i=\hat{t}+1}^{\hat{t}+\hat{k}}\hat{g}_{i}(\hat{x})\hat{g}_{i}^{*}(\hat{x}) \right)$$ 
be the factorization of $\hat{x}^{\frac{2n}{\ell}}-1$ into irreducible factors. Let $$\Omega: \fq[\hat{x}]/(\hat{x}^{\frac{2n}{\ell}}-1) \hookrightarrow \fqQ, \hat{x} \mapsto x^\ell$$ be the embedding of $\fq[\hat{x}]/(\hat{x}^{\frac{2n}{\ell}}-1)$ into $\fqQ$. In analogy with \eqref{eq:genidem} one can define the idempotent $\hat{e}_{h}(\hat{x})$ for a monic polynomial $h \in \fq[\hat{x}]/(\hat{x}^{\frac{2n}{\ell}}-1)$ which divides $\hat{x}^{\frac{2n}{\ell}}-1$. Moreover, using \eqref{eq:idemprop}, it is easy to see that if $h$ is irreducible one can write
\begin{align*}
    \hat{e}_{h}(x^{\ell})&=\left( \sum_{\substack{j=1 \\ f_{j}(x) \mid h(x^\ell)}}^{r+s}e_{f_{j}}(x) \right) + \left( \sum_{\substack{j=r+1 \\ f_{j}^{*}(x) \mid h(x^\ell)}}^{r+s}e_{f_{j}^{*}}(x) \right)
    +\left( \sum_{\substack{i=1 \\ g_{i}(x) \mid h(x^\ell)}}^{k+t}e_{g_{i}}(x) \right)+\left( \sum_{\substack{i=t+1 \\ g_{i}^{*}(x) \mid h(x^\ell)}}^{k+t}e_{g_{i}^{*}}(x) \right)
\end{align*}
\normalsize
The following observations on generating idempotents immediately follow
\begin{lemma} \label{lem:idem}
Let $(\hat{f}_{j})_{1 \le j \le r+s}, (\hat{f}_{j}^{*})_{r+1 \le j \le r+s}, (\hat{g}_{i})_{1 \le i \le t+k}$ and $(\hat{g}_{i}^{*})_{t+1 \le i \le t+k}$ be the irreducible factors of $\hat{x}^{\frac{2n}{\ell}}-1$ in $\fq[\hat{x}]/(\hat{x}^{\frac{2n}{\ell}}-1)$. 
    \begin{itemize}
    \item For $1 \le j \le \hat{r}$
    \begin{align*}
        \hat{e}_{\hat{f}_{j}}(x^{\ell})&= \left( \sum_{\substack{i=1 \\ f_{i}(x) \mid \hat{f}_{j}(x^\ell)}}^{r}e_{f_{i}}(x) \right) + \left( \sum_{\substack{i=r+1 \\ f_{i}(x) \mid \hat{f}_{j}(x^\ell)}}^{r+s}(e_{f_{i}}(x)+ e_{f_{i}^{*}}(x))\right)\\
        & \hspace{10em}+\left( \sum_{\substack{i=1 \\ g_{i}(x) \mid \hat{f}_{j}(x^\ell)}}^{t}e_{g_{i}}(x) \right)
    +\left( \sum_{\substack{i=t+1 \\ g_{i}(x) \mid \hat{f}_{j}(x^\ell)}}^{t+k}(e_{g_{i}}(x)+e_{g_{i}^{*}}(x)) \right).
    \end{align*}
    \item For $\hat{r}+1 \le j \le \hat{r}+\hat{s}$

    \smaller
    \begin{align*}
        \hat{e}_{\hat{f}_{j}}(x^{\ell})= \left( \sum_{\substack{i=r+1 \\ f_{i}(x) \mid \hat{f}_{j}(x^\ell)}}^{r+s}e_{f_{i}}(x) \right) + \left( \sum_{\substack{i=r+1 \\ f_{i}^{*}(x) \mid \hat{f}_{j}(x^\ell)}}^{r+s}e_{f_{i}^{*}}(x)\right)
      +\left( \sum_{\substack{i=t+1 \\ g_{i}(x) \mid \hat{f}_{j}(x^\ell)}}^{t+k}e_{g_{i}}(x) \right)
    +\left( \sum_{\substack{i=t+1 \\ g_{i}^{*}(x) \mid \hat{f}_{j}(x^\ell)}}^{t+k}e_{g_{i}^{*}}(x) \right),\\
      \hat{e}_{\hat{f}_{j}^{*}}(x^{\ell})= \left( \sum_{\substack{i=r+1 \\ f_{i}(x) \mid \hat{f}_{j}(x^\ell)}}^{r+s}e_{f_{i}^{*}}(x) \right) + \left( \sum_{\substack{i=r+1 \\ f_{i}^{*}(x) \mid \hat{f}_{j}(x^\ell)}}^{r+s}e_{f_{i}}(x)\right)
      +\left( \sum_{\substack{i=t+1 \\ g_{i}(x) \mid \hat{f}_{j}(x^\ell)}}^{t+k}e_{g_{i}^{*}}(x) \right) +\left( \sum_{\substack{i=t+1 \\ g_{i}^{*}(x) \mid \hat{f}_{j}(x^\ell)}}^{t+k}e_{g_{i}}(x) \right).
    \end{align*}
   \item \normalsize For $1 \le i \le \hat{t}$
    \begin{align*}
        \hat{e}_{\hat{g}_{i}}(x^{\ell})&= \left( \sum_{\substack{j=1 \\ f_{j}(x) \mid \hat{g}_{i}(x^\ell)}}^{r}e_{f_{j}}(x) \right)
    +\left( \sum_{\substack{j=r+1 \\ f_{j}(x) \mid \hat{g}_{i}(x^\ell)}}^{r+s}(e_{f_{j}}(x)+e_{f_{j}^{*}}(x)) \right)\\
        & \hspace{10em}+\left( \sum_{\substack{j=1 \\ g_{j}(x) \mid \hat{g}_{i}(x^\ell)}}^{t}e_{g_{j}}(x) \right) + \left( \sum_{\substack{j=t+1 \\ g_{j}(x) \mid \hat{g}_{i}(x^\ell)}}^{t+k}(e_{g_{j}}(x)+ e_{g_{j}^{*}}(x))\right).
    \end{align*}
    \item For $\hat{t}+1 \le i \le \hat{t}+\hat{k}$
    \smaller{\begin{align*}
        \hat{e}_{\hat{g}_{i}}(x^{\ell})&= \left( \sum_{\substack{j=r+1 \\ f_{j}(x) \mid \hat{g}_{i}(x^\ell)}}^{r+s}e_{f_{j}}(x) \right) + \left( \sum_{\substack{j=r+1 \\ f_{j}^{*}(x) \mid \hat{g}_{i}(x^\ell)}}^{r+s}e_{f_{j}^{*}}(x)\right)
      +\left( \sum_{\substack{j=t+1 \\ g_{j}(x) \mid \hat{g}_{i}(x^\ell)}}^{t+k}e_{g_{j}}(x) \right)
    +\left( \sum_{\substack{i=t+1 \\ g_{j}^{*}(x) \mid \hat{g}_{i}(x^\ell)}}^{t+k}e_{g_{j}^{*}}(x) \right),\\
      \hat{e}_{\hat{g}_{i}^{*}}(x^{\ell})&= \left( \sum_{\substack{j=r+1 \\ f_{j}(x) \mid \hat{g}_{i}^{*}(x^\ell)}}^{r+s}e_{f_{j}^{*}}(x) \right) + \left( \sum_{\substack{j=r+1 \\ f_{j}^{*}(x) \mid \hat{g}_{i}^{*}(x^\ell)}}^{r+s}e_{f_{j}}(x)\right)
      +\left( \sum_{\substack{j=t+1 \\ g_{j}(x) \mid \hat{g}_{i}^{*}(x^\ell)}}^{t+k}e_{g_{j}^{*}}(x) \right) +\left( \sum_{\substack{j=t+1 \\ g_{j}^{*}(x) \mid  \hat{g}_{i}^{*}(x^\ell)}}^{t+k}e_{g_{j}}(x) \right).
    \end{align*}}
\end{itemize}
\end{lemma}
\normalsize
\begin{theorem}
    Set $\hat{C}_{h}=(h)$ and let $C=\fqQ \Omega(\hat{C_{h}})$ be the code induced by $\hat{C}_{h}$. Then $C$ decomposes as
$$C \cong \bigoplus_{j=1}^{r+s}C_{j} \oplus \bigoplus_{i=1}^{t+k}C'_{i},$$ where
$$C_{j}=\begin{cases}
A_j\quad & j \in S_1, \\
  I_{j}(0,1) \quad & j \in S_2, \\
  I_{j}(1,0) \quad & j \in S_3, \\
   0 \quad & j \notin S_1 \cup S_2 \cup S_3,
\end{cases}\quad C'_{i}=\begin{cases}
B_i\quad & i \in S_4, \\
  J_{i}(0,1) \quad & i \in S_5, \\
  J_{i}(1,0) \quad & i \in S_6, \\
   0 \quad & i \notin S_4 \cup S_5 \cup S_6,
\end{cases}$$
and
\begin{align*}
    S_1 &= \{j \in \{1, \dots, r+s\}: f_j(x) \nmid h(x^{\ell}) \wedge f_{j}^{*}(x) \nmid h(x^{\ell}) \},\\
      S_2 &= \{j \in \{\delta(n)+1, \dots, r+s\}: f_j(x) \nmid h(x^{\ell}) \wedge f_{j}^{*}(x) \mid h(x^{\ell}) \},\\
        S_3 &= \{j \in \{\delta(n)+1, \dots, r+s\}: f_j(x) \mid h(x^{\ell}) \wedge f_{j}^{*}(x) \nmid h(x^{\ell}) \},\\
          S_4 &= \{i \in \{1, \dots, k+t\}: g_i(x) \nmid h(x^{\ell}) \wedge g_{i}^{*}(x) \nmid h(x^{\ell}) \},\\
            S_5 &= \{i \in \{\mu(n)+1, \dots, k+t\}: g_i(x) \nmid h(x^{\ell}) \wedge g_{i}^{*}(x) \mid h(x^{\ell}) \},\\
              S_6 &= \{i \in \{\mu(n)+1, \dots, k+t\}: g_i(x) \mid h(x^{\ell}) \wedge g_{i}^{*}(x) \nmid h(x^{\ell}) \}.
\end{align*}
\begin{proof}
Recalling the generating idempotent $e_{g}(x)$ given in \eqref{eq:genidem} and that for $g \in \fq[x]/(x^{2n}-1)$, a monic divisor of $x^{2n}-1$ with $g=g_{1} \cdot g_{2}$, it holds $e_{g}(x)=e_{g_{1}}(x)+e_{g_{2}}(x)$,
the generating idempotent of the code \( \hat{C_{h}} \) has the form:
\[
\hat{e}_{\hat{C_{h}}}(\hat{x})=
\left( \sum_{\substack{j=1 \\ \hat{f}_{j}(\hat{x}) \nmid h(\hat{x})}}^{r+s} \hat{e}_{\hat{f}_{j}}(\hat{x}) \right) 
+ \left( \sum_{\substack{j=1 \\ \hat{f}_{j}^{*}(\hat{x}) \nmid h(\hat{x})}}^{r+s} \hat{e}_{\hat{f}_{j}^{*}}(\hat{x}) \right)
+ \left( \sum_{\substack{i=1 \\ \hat{g}_{i}(\hat{x}) \nmid h(\hat{x})}}^{t+k} \hat{e}_{\hat{g}_{i}}(\hat{x}) \right)
+ \left( \sum_{\substack{i=1 \\ \hat{g}_{i}^{*}(\hat{x}) \nmid h(\hat{x})}}^{t+k} \hat{e}_{\hat{g}_{i}^{*}}(\hat{x}) \right).
\]

This gives:
\[
C = \fqQ \Omega(\hat{C_{h}}) = \fqQ \Omega\left( \fq[\hat{x}]/(\hat{x}^{\frac{2n}{\ell}}-1)e_{\hat{C_{h}}}(\hat{x}) \right)
= \fqQ e_{\hat{C_{h}}}(x^\ell)
\]
and from Lemma \ref{lem:idem} we have
\[ \displaystyle
\hat{e}_{\hat{C}_{h}}(x^\ell) =
\left( \sum_{\substack{j=1 \\ f_{j}(x) \nmid h(x^\ell)}}^{r+s} e_{f_{j}}(x) \right)
+ \left( \sum_{\substack{j=r+1 \\ f_{j}^{*}(x) \nmid h(x^\ell)}}^{r+s} e_{f_{j}^{*}}(x) \right)
+ \left( \sum_{\substack{i=1 \\ g_{i}(x) \nmid h(x^\ell)}}^{t+k} e_{g_{i}}(x) \right)
+ \left( \sum_{\substack{i=t+1 \\ g_{i}^{*}(x) \nmid h(x^\ell)}}^{t+k} e_{g_{i}^{*}}(x) \right).
\]
        Applying Theorem \ref{thm:main} the statement follows.
\end{proof}
\end{theorem}
\section{Conclusion}
The block components $(C_{j})_{1 \le j \le r+s}, (C_{i}')_{1 \le i \le t+k}$ of a code $C \subseteq \fqQ$ with decomposition $C\cong \bigoplus_{j=1}^{r+s}C_j \oplus \bigoplus_{i=1}^{k+t}C'_{i}$ were identified via its generating idempotent. Afterwards this result is applied to establish the structure of codes induced
by cyclic groups. Understanding the structure and decomposition of such induced codes is a key
step for the construction of lifted product codes as demonstrated in \cite{willenborg2023dihedral}.
Similarly as one can determine the blocks of codes induced by cyclic group codes one can also obtain the blocks of codes in $\fqQ$ induced by dihedral group codes, see \cite{willenborg2023dihedral}. However this case requires a rigourous computation of the generating idempotent of dihedral codes, see \cite{vedenev2020relationship}, resulting in a more detailed analysis, see \cite{gao2021idempotents, brochero2022wedderburn}. As already demonstrated in \cite{vedenev2020codes} where the bases, the generating and the check matrices of the dihedral codes are constructed one can use the algebra homomorphisms $(\tau_{j})_{1 \le j \le r+s}, (\eta_{i})_{1 \le i \le t+k}$ explicitly given in \cite{gao2021idempotents}, to extend this to the generalized quaternion group algebra $\fqQ.$ 
\section*{Acknowledgements}
This work is part of my PhD thesis under the supervision of Anna-Lena Horlemann. I would like to thank Anna-Lena Horlemann for her valuable guidance, constant encouragement to work on group codes and for helpful comments on my work.

\bibliographystyle{amsplain}
 \bibliography{biblio_1.bib}

\end{document}